\documentclass[12pt, oneside,a4page]{article}

\usepackage{graphicx}

\begin{document}


\begin{center}
{\large\bf Dynamic Behaviors of Supersonic Granular Media under
Vertical Vibration} \\

{Kai Huang \footnote{Corresponding author. Tel. +86 25 8359 4184;
Fax: +86 25 8331 5557; Postal address: P.O.Box 1004, Nanjing
Univeristy, Nanjing, China 210093; Email address:
huangkai1996@nju.org.cn\\}, Peng Zhang, Guoqing Miao
and Rongjue Wei}\\

{\it The State Key Laboratory of Modern Acoustics and The
Institute of Acoustics, Nanjing University, Nanjing 210093} \\
\end{center}

\parbox[c]{1\textwidth}{ {\bf Abstract} \\
\small \indent We present experimental study of vibrofluidized
granular materials by high speed photography. Statistical results
present the averaged dynamic behaviors of granular materials in
one cycle, including the variations of height, velocity and
mechanical energy of the center of mass. Furthermore, time-space
distribution of granular temperature which corresponds to the
random kinetic energy shows that a temperature peak forms in the
compression period and propagates upward with a steepened front.
The Mach number in the steepened front is found to be greater than
unity, indicating a shock propagating in the supersonic granular
media.
}

\vspace{1cm}
\parindent 0.5cm

\section{Introduction}
Granular materials are conglomerations of macroscopic particles.
They are so ubiquitous that they attract interest from people all
over the world for a long time. It has been known that with
different energy input, granular materials behave like solids,
liquids or gases~\cite{Rev1}. When fluidized, granular materials
show many interesting phenomena such as surface
pattern~\cite{Pattern}, convection~\cite{Conv}, heap
formation~\cite{Heap}, localized excitation~\cite{Oscillon} and so
on. To describe dynamical behaviors of fluidized granular
materials, properties such as density, pressure and temperature
have been defined and various theories have been
introduced~\cite{Jenkins}. Correspondingly noninvasive methods for
detecting these properties from statistical average of
experimental data is created~\cite{Method1}. Moreover computer
simulations~\cite{Simu1} are performed to compare with
experimental results. Recent simulation results~\cite{shock}
indicate that it is relatively easy for vertical vibrated granular
materials to be supersonic. In this paper, we use high speed
photography to explore time-dependent behaviors of a two
dimensional vibrofluidized granular materials and give evidence of
supersonic granular flow by experimental results of granular
temperature waves.

%



\section{Method}
The experiment is conducted with a rectangular container mounted
on the vibration exciter (Br\"{u}el \& Kj$\ae$r 4805), which is
controlled by a function generator (type HP 3314A). We use steel
spheres with diameter $d=7mm$ and density $\rho=7900kg/m^3$. The
container is made up of two parallel
$155mm$(length)$\times$$280mm$(height) glass plates separated by
$7.5mm$ vertically adhered in a Plexiglas bracket. Particle number
$N$ changes from $60$ to $210$. The vibrator undergoes sinusoidal
vibrations controlled by driving parameters frequency $f$ and
nondimensional acceleration $\Gamma = 4 \pi^{2} f^{2}A/g$, where
$A$ is the driving amplitude, and $g$ the gravitational
acceleration. High speed camera (Redlake MASD MotionScope PCI
2000sc) is employed to record the movements of all spheres. A
frequency multiplication and phase lock circuit is used to
generate external trigger signals for the camera. The acquisition
rate is $N_pf$ with the multiple number $N_p=25$.

Image processing technique \cite{Method1} is used to track
locations and velocities of particles. Every recorded image
($22d$(wide)$\times$$30d$(height)) is divided uniformly into
$10$(wide)$\times$$30$(height) cells. The cells are numbered by
indices $i$ and $j$ which specify the positions of the cells along
the horizontal and vertical directions. To obtain time-dependant
behaviors of granular materials, average is performed at $N_p$
phase points over all the $326$ cycles captured. Granular
temperature is designated by
$T(i,j,k)=\sum^{N_c}_{n=1}\frac{1}{2}\vert {\bf v}_n-{\bf
v}_b(i,j,k) \vert^2/N_c$, in which $N_c$ is total number of
particles in cell ($i,j$) at time $k$, ${\bf v}_n$ is velocity
vector of the $nth$ particle, and background velocity ${\bf
v}_b(i,j,k)$ is the mean of the particle velocities in cell
($i$,$j$) at the same time $k$ for all the cycles.


\section{Results and Discussion}
\begin{figure}
\begin{center}
    \resizebox{0.55\textwidth}{!}{\includegraphics{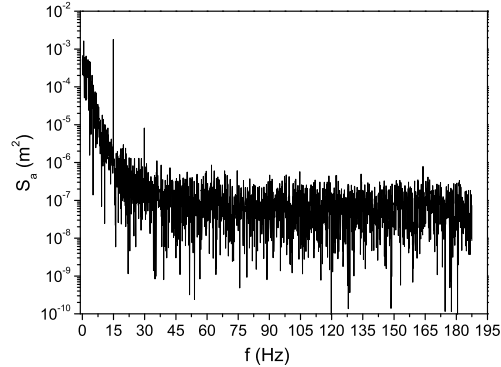}}
\end{center}
\caption{\label{fig:spec} Experimental results of $S_a$ (squared
amplitude spectrum of c.m.) with driving parameter $\Gamma=5$ and
$f=15Hz$. The acquisition rate is $375$ images per second.}
\end{figure}

With the increase of $\Gamma$ from a critical value and other
parameters ($f=15Hz$, $N=60$) fixed, granular materials become
fluidized from upper to lower layers. The packing fraction
decreases and particles change locations relative to their
neighbors frequently. The squared amplitude spectrum of the
vertical displacement of the center of mass(c.m. is used as an
abbreviation for the center of mass hereafter) in laboratory
reference frame with $\Gamma=5$ is shown in Fig.~\ref{fig:spec}.
It is composed of a noisy background and two lines at the basic
and second harmonic. The spectrum line at the basic harmonic is
about $3$ order of magnitude larger than that at the second
harmonic. This indicates that granular materials as a whole move
periodically with the vibrating frequency of the plate. Although
spectrum lines at higher harmonic begin to surpass the background
noise and become comparable with that of the second harmonic with
the increase of $N$ from $60$ to $150$ and other parameters fixed,
the periodicity still exists for the spectrum line at the basic
frequency keeps one order larger than that at higher harmonic. The
periodicity also exists with other driving parameters we used.

\begin{figure}
\begin{center}
    \resizebox{0.45\textwidth}{!}{\includegraphics{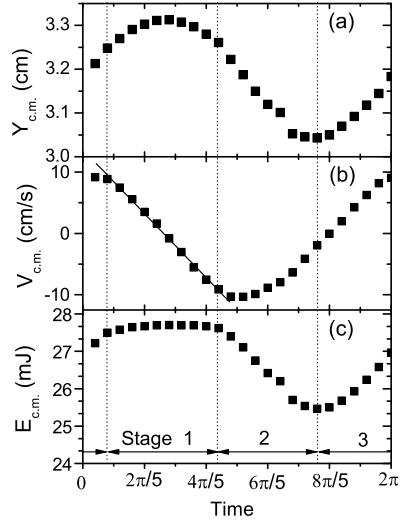}}
\end{center}
\caption{\label{fig:mcy} Height(a), velocity(b) and mechanical
velocity(c) of the center of mass as a function of time with
driving parameters $\Gamma=5$ and $f=15Hz$, time $0$ corresponds
to the time when the plate is at its maximum height.}
\end{figure}

Fig.~\ref{fig:mcy} shows time-dependant behaviors of the center of
mass, which indicates that granular materials as a whole undergo
three stages in one cycle. In the first stage (from $4\pi/25$ to
$22\pi/25$), granular materials fly freely. The trajectory of the
center of mass is parabolic, its velocity decreases almost
linearly with time and the mechanical energy (defined as
$E_{c.m.}=NmgY_{c.m.}+NmV_{c.m.}^2/2$, in which $m$ is the mass of
one particle) fixes at about $27.5mJ$. In Stage $2$ (from
$22\pi/25$ to $38\pi/25$), granular materials collide with the
plate and compress. Height of the center of mass continues to
decrease but deviates from originally parabolic track. Particles
at lower layers change their moving direction by colliding with
the plate which leads to the increase of negative c.m. velocity.
In this stage the mechanical energy of c.m. drastically decreases,
which is due to the decrease of both the height and scalar
velocity. The third stage is the restoration stage. It starts at
$38\pi/25$ when granular materials reach the most compact state.
After this time c.m. of granular materials changes its direction
and moves upward and all of the $Y_{c.m.}$, $V_{c.m.}$ and
$E_{c.m.}$ increase. This stage lasts until $4\pi/25$ at the next
vibration cycle when granular layer begins to leave the plate and
fly freely again. Then a new cycle begins. Experiments with
different driving parameters ($f=15Hz$ and $20Hz$, $\Gamma=2$ to
$8$) and particle numbers ($N=60$ to $210$) are also performed and
similar time-dependent behaviors of the center of mass are
obtained except for some parameters with which the well of input
energy exists \cite{miao}.

\begin{figure}
\begin{center}
    \resizebox{0.45\textwidth}{!}{\includegraphics{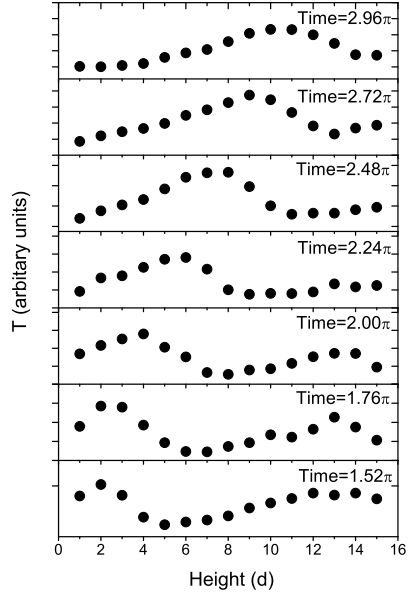}}
\end{center}
\caption{\label{fig:shock}Granular temperature as a function of
height from $Time=1.52\pi$ to $Time=2.96\pi$ of the next cycle
with parameters the same as those in Fig.~\ref{fig:mcy}.}
\end{figure}

Vibrofluidized granular materials are strongly dissipative for
inelastic collisions between particles. The vibrating plate who
acts as energy source to keep granular media fluidized are also
the source of mechanical waves propagating in the media.
Fig.~\ref{fig:shock} shows the propagation of a temperature wave
in one vibration cycle. As granular materials fly freely and
expands, the granular temperature as a whole decreases by the
dissipation. In the second stage, as an obstacle the plate collide
with granular materials. The velocities of particles at the lower
layers drop to that of the plate while particles at higher layers
still fly downward freely. Thus at $Time=1.52\pi$ an active area
of collisions between particles as well as a temperature peak is
formed. Between $Time=1.52\pi$ and $1.76\pi$, the value of the
temperature peak increases to a peak value $0.097mJ$ at about
$1.76\pi$ and then slowly decreases as time elapses. After
$Time=1.76\pi$, the perturbation of granular temperature
propagates upward with a uniform velocity $v_T=1.51m/s$. At
$Time=2.96\pi$ of the next cycle, the wave transmit to height $11$
as its amplitude decays. With the granular temperature and volume
fraction obtained by experiments, we calculate the wave
propagation velocity and the Mach
number~\cite{shock,acousticSpeed}. The average phase velocity
($c=0.438m/s$) is comparable with the granular flow velocity
relative to the vibrating plate. The Mach number increases to be
greater than unity at the steepened temperature front when
granular layer collides with the plate. At the end time of the
compression period $Time=1.52\pi$ the Mach number reaches its
maximum $1.71$, which indicates that there exist supersonic
granular flow as the collisions between particles and the plate
occur.

%
%
%

\section{Conclusions}
In summary, we present experimental study of supersonic granular
materials. Power spectrum of the height of the center of mass
indicates that granular materials as a whole move periodically
with the driving frequency. Statistical averages of the height,
velocity and mechanical energy of the center of mass in one
vibration cycle indicate that granular materials undergo three
stages: free flight stage, compression stage and restoration
stage. Time-space distribution of granular temperature shows that
a temperature peak forms within the compression period of the
granular materials and propagates upward with a steepened front
when the particles collide with the plate and compress. The
existence of this steepened temperature front and the maximum Mach
number in this media demonstrate that it is easy for vertical
vibrated granular flow to be supersonic. This is due to that the
acoustic velocity in granular materials is less than that of
granular flow relative to the vibrating plate.

Internal wave propagations of mechanical waves take the role of
transporting energy from the plate to granular layer. Study of
shock propagation in vibrofluidized granular materials and its
relationship with surface instability will be the subject of the
future work, which may provide useful information about the
mechanism of surface waves and other
interesting phenomena of vibrofluidized granular materials.\\


\noindent {\large \bf Acknowledgement}\\
This work was supported by the Special Funds for Major State Basic
Research Projects, National Natural Science Foundation of China
through Grant No. 10474045 and No. 10074032, and by the Research
Fund for the Doctoral Program of Higher Education of China under
Grant No. 20040284034.

%
%
%
%
%
%
%
%
%
%
%
%
%
%
%


\begin{thebibliography}{0}
\bibitem{Rev1}
H.M. Jaeger, S.R. Nagel, Rev. Mod. Phys. 68 (1996) 1259-1273.
\bibitem{Pattern}
F. Melo, P. Umbanhowar, H.L. Swinney, Phys. Rev. Lett. 72 (1994)
172-175; F. Melo, P.B. Umbanhowar, H.L. Swinney, Phys. Rev. Lett.
75 (1995) 3838-3841.
\bibitem{Conv}
J.B. Knight, E.E. Ehrichs, V.Y. Kuperman, J.K. Flint, H.M. Jaeger,
S.R. Nagel, Phys. Rev. {\bf E} 54 (1996) 5726-5738; G.Q. Miao, K.
Huang, Y. Yun, R. J. Wei, Euro. Phys. J. {\bf B} 40 (2004)
301-304.
\bibitem{Heap} P. Evesque, J. Rajchenbach, Phys. Rev. Lett. 62 (1989) 44-46; K. Huang, G. Q. Miao, R. J. Wei, Int. J. Mod. Phys. {\bf B} 17
(2003) 4222-4226.
\bibitem{Oscillon} P.B. Umbanhowar, F. Melo, H.L. Swinney, Nature 382 (1996) 793-796.
\bibitem{Jenkins} J. T. Jenkins and S. B. Savage, J. Fluid Mech.
130 (1983) 187-202; P. K. Haff, J. Fluid Mech. 134 (1983) 401-430.
\bibitem{Method1} S. Warr, G.T.H. Jacques, J.M. Huntley, Powder Technol. 81 (1994)
41-56.
\bibitem{Simu1} C. Bizon, M.D. Shattuck, J.B. Swift, W.D.
McCormick, H.L. Swinney, Phys. Rev. Lett. 80 (1998)57-60.
\bibitem{shock} J.Bougie, S.J. Moon,
J.B. Swift, H.L. Swinney, Phys. Rev. {\bf E} 66 (2002) 051301.
\bibitem{miao} G.Q. Miao, L.Sui, R.J. Wei, Phys. Rev. {\bf E} 63 (2001) 031304.
\bibitem{acousticSpeed}
S.B. Savage, J. Fluid Mech. 194 (1988) 457-478.
\end{thebibliography}
\end{document}